\documentclass[a4paper]{jpconf}
\usepackage{epsfig}
\usepackage{axodraw}
\usepackage{graphicx}
\usepackage{bm}
\usepackage{setspace}
\usepackage{color}

\newcommand{\la}{{\lambda}}
\newcommand{\ie}{{\it i.e.}}
\newcommand{\eg}{{\it e.g.}}

\newcommand{\be}{\begin{equation}}
\newcommand{\ee}{\end{equation}}
\newcommand{\beq}{\begin{equation}}
\newcommand{\eeq}{\end{equation}}
\newcommand{\bea}{\begin{eqnarray}}
\newcommand{\eea}{\end{eqnarray}}
\newcommand{\er}{\end{eqnarray}}
\newcommand{\ba}{\begin{array}}
\newcommand{\ea}{\end{array}}
\newcommand{\ei}{\end{itemize}}
\newcommand{\bn}{\begin{enumerate}}
\newcommand{\en}{\end{enumerate}}
\newcommand{\bc}{\begin{center}}
\newcommand{\ec}{\end{center}}

\def\bY{{\bf Y}}

\def\bkappa{\bm{\kappa}}
\def\bk5{\bm{\kappa}_5}
\def\bD{\bm{\Delta}}
\def\bA{{\bf A}}
\def\bB{{\bf B}}
\def\bC{{\bf C}}

\def\bU{{\bf U}}
\def\bV{{\bf V}}
\def\bmm{{\bf m}}

\def\bk{{\cal {\bf K}}}
\def\tl{{\tilde{L}}}

\def\tm{{\tilde{m}}}
\def\te{{\tilde{e^c}}}

\def\unity{{\hbox{1\kern-.6mm l}}}

\newcommand{\no}{\nonumber}

\newcommand{\BR}{{\rm BR}}

\newcommand{\gev}{\mbox{GeV}}

\newcommand{\ga}{\gamma}

\newcommand{\gsim}{\lower.7ex\hbox{$\;\stackrel{\textstyle>}{\sim}\;$}}

\begin{document} \title{A common source for neutrino and sparticle
masses\footnote{Based on Ref.~\cite{Brignole:2010nh}. Presented at
``PASCOS 2010, the 16th International Symposium on Particles,
Strings and Cosmology'' (Valencia, Spain) and ``SUSY 2010, the
18th International Conference on Supersymmetry and Unification of
Fundamental Interactions'' (Bonn, Germany).}}

\author{Andrea Brignole}

\address{INFN, Sezione di Padova, I-35131 Padua, Italy}

\author{Filipe R. Joaquim}

\address{CERN, Theory Division, CH-1211 Geneva 23, Switzerland}

\author{Anna Rossi}

\address{Dipartimento di Fisica ``G.~Galilei'', Universit\`a di
Padova, I-35131 Padua, Italy}

\begin{abstract}
We discuss supersymmetric scenarios in which neutrino masses
arise from effective $d=6$ operators in the K\"ahler potential
(including SUSY-breaking insertions). Simple explicit
realizations of those K\"ahler operators are presented in the context of the type II seesaw.
An appealing scenario emerges upon identifying the seesaw mediators with SUSY-breaking messengers.
\end{abstract}

The smallness of neutrino masses can be explained by the conventional seesaw mechanism in which $m_\nu \sim
v^2 /M$, where $v$ is the electroweak scale and $M \gg v$ is a
heavy mass. It is also conceivable that neutrino masses are
suppressed by a higher power of the heavy scale $M$, like $m_\nu \sim m v^2/M^2$,
where $m\ll M$ is another mass parameter. In a SUSY framework
this mass behaviour may stem from either
superpotential or K\"{a}hler $d=6$, $\Delta L =2$ effective
operators. Two examples of the latter have been proposed
in \cite{CEN}, namely $\int  \! d^4 \theta \, (H^\dagger_1L)
(H_2 L)/M^2$ and $\int \!  d^4 \theta \, (H^\dagger_1 L)^2/M^2$,
which imply $m \sim \mu$ (the superpotential Higgsino mass parameter). We
point out the importance of including SUSY-breaking insertions to this approach and
find new contributions to neutrino masses.
We also present the simplest explicit realization of the
K\"{a}hler operator $(H^\dagger_1 L)^2$, including SUSY-breaking
effects, and discuss a predictive scenario where the heavy seesaw mediators
are also messengers of SUSY breaking.

\section{Neutrino masses from K\"ahler operators
and broken SUSY}
\vspace*{0.2cm}
Let us focus on the $\Delta L=2$ effective operator $(H^\dagger_1 L)^2/M^2$.
In general, this will appear along with similar operators containing insertions
of the type $X/M_S$, ${X^\dagger}/{M_S}$,
$ X X^\dagger/{M_S^2}$, where $X=\theta^2 F_X$ is a SUSY-breaking
spurion superfield and $M_S$ is the
SUSY-breaking mediation scale, which can be either larger or smaller
than $M$.  Hence, the relevant $\Delta L=2$
effective lagrangian can be parametrized as
\be
\label{K2}
{\cal L}_{\rm
eff}  = \int  \!  d^4 \theta \frac{1}{2 M^2} \left( \bkappa +
\theta^2 \bB_\kappa + \bar{\theta}^2 \tilde{\bB}_\kappa + \theta^2
\bar{\theta}^2 \bC_\kappa \right)_{ij} (H^\dagger_1 L_i)
(H^\dagger_1 L_j) \,\, + {\rm h.c.}\,,
\ee
where $\bkappa$ is dimensionless, $\bB_\kappa$, $\tilde{\bB}_\kappa$
and $\bC_\kappa$ are dimensionful SUSY-breaking parameters and
$i,j=e,\mu,\tau$ are flavour indices. The magnitude and flavour structure
of all those parameters depend on the underlying physics which
generates the effective operators.

The scale dependence of the above quantities is governed by their
renormalization group equations (RGEs),
which can be derived using the general expression of the
one-loop corrected  K\"ahler potential obtained in \cite{ab}.
The RGE for $\bkappa$ can be found in~\cite{CEN}, while the ones
for the remaining operator coefficients have been obtained in~\cite{Brignole:2010nh}.
For instance, the RGE for $\tilde{\bB}_\kappa$ is:
\bea
\label{rgebkt} 8 \pi^2  \frac{d {\tilde \bB}_\kappa}{dt } & = &
\left[ g^2 + g'^2 +  {\rm Tr}( \bY_e^\dagger \bY_e + 3
\bY^\dagger_d \bY_d)  \right] {\tilde \bB}_\kappa   -
\frac{1}{2}\left[ {\tilde \bB}_\kappa \bY^\dagger_e \bY_e +
(\bY^\dagger_e \bY_e )^T {\tilde \bB}_\kappa \right]
\nonumber \\
& &  \!\!\! + \left[ g^2 M^*_2 + g'^2 M^*_1 - 2 \, {\rm Tr}(
\bA_e^\dagger \bY_e + 3 \bA^\dagger_d  \bY_d)  \right] \bkappa +
\bkappa \, \bA^\dagger_e \bY_e + (\bA^\dagger_e \bY_e )^T
\bkappa\,.\eea
%

%%%%%%%%%%%%%%%%%%%%%%%%%%%%%%%%%%%%%%%%%%%%%%%%%%%%%%%
All four  operators of eq.~(\ref{K2}) contribute to neutrino masses.
Those  with coefficients $\bkappa$ and $\tilde{\bB}_\kappa$
contribute  to $\bmm_\nu$ at the tree-level,
in such a way that
$
\bmm_\nu \simeq \bmm^{(\kappa)}_\nu +
\bmm^{(\tilde{B}_\kappa)}_\nu\,.
$
The $\bkappa$-operator leads to a lagrangian term of the form $
(F_{H_1}^\dagger L)(H_1^\dagger L)$, which contributes to neutrino masses as~\cite{CEN}
\be
\label{mnuk}
\bmm^{(\kappa)}_\nu = 2\,  \bkappa \; \mu \, \frac{ v^2}{M^2}
\sin\beta \cos\beta\,.
\ee
The $\tilde{\bB}_\kappa$-operator gives a
lagrangian term of the form $ (H_1^\dagger L)^2$, which induces~\cite{Brignole:2010nh}
\be
\label{mnubtk}
\bmm^{(\tilde{B}_\kappa)}_\nu =
\tilde{\bB}_\kappa \frac{v^2}{M^2} \cos^2\beta\,.
\ee
This novel contribution to $\bmm_\nu$ can dominate over $\bmm^{(\kappa)}_\nu$.
The remaining operators of eq.~(\ref{K2}), with coefficients $\bB_\kappa$ and
$\bC_\kappa$, contribute to neutrino masses through finite one-loop diagrams
involving gauginos and sleptons.
Considering the soft mass matrix
of `left-handed' sleptons  $\tilde{L}$ as $\bmm^2_{\tl} = {\tilde
m}_L^2 ( \unity + \bD_L)$ (${\tilde m}_L^2$ sets the overall
mass scale and the dimensionless matrix $\bD_L$ accounts for the
flavour dependence), we get the following contributions to $\bmm_\nu$
(at first order in $\bD_L$)
\bea \label{deltambk} \scriptstyle\delta_{B_\kappa} \bmm_\nu  \scriptstyle& \!\!\!\!\scriptstyle\simeq &\!\!\!\!\scriptstyle
\frac{1}{32 \pi^2} \left[ - \left( \frac{g^2}{M_2} \, f_{L2}
+\frac{g'^2}{M_1} \, f_{L1} \right) \bB_\kappa  + \left( \frac{g^2}{M_2} \, h_{L2} +\frac{g'^2}{M_1} \,
h_{L1} \right) (\bB_\kappa  \, \bD_L + \bD_L^T \, \bB_\kappa )
\right] 2 \mu \frac{v^2}{M^2} \sin\beta \cos\beta\,,
\\
\label{deltamck} \scriptstyle\delta_{C_\kappa} \bmm_\nu \scriptstyle & \!\!\!\!\scriptstyle\simeq &\!\!\!\!\scriptstyle
\frac{1}{32 \pi^2} \left[ - \left( \frac{g^2}{M_2} \, f_{L2}
+\frac{g'^2}{M_1} \, f_{L1} \right) \bC_\kappa   + \left( \frac{g^2}{M_2} \, h_{L2} +\frac{g'^2}{M_1} \,
h_{L1} \right) (\bC_\kappa  \, \bD_L + \bD_L^T \, \bC_\kappa )
\right] \frac{v^2}{M^2} \cos^2 \! \beta\,, \eea
where $f_{L a}= f
({\tilde m}_L^2/|M_a|^2)$, $h_{L a}= h({\tilde m}_L^2/|M_a|^2)$,
$f(x)=(x-1- \log x)/(x-1)^2$ and
$h(x)=(x^2-1-2x \log x)/(x-1)^3$. Both the flavour structure and the size of
$\delta_{B_\kappa}\bmm_\nu$, $\delta_{C_\kappa} \bmm_\nu$ are
model dependent.

%%%%%%%%%%%%%%%%%%%%%%%%%%%%%%%%%%%%%%%%%%%%%%%%%%%%%%%

\section{Type II seesaw realizations}
\vspace*{0.2cm}
Among the three variants of the seesaw
mechanism (which generate the familiar $d=5$ superpotential
operator at the tree level), the type
II is the natural one in which the above $d=6$ operators
emerge. In fact, the tree-level exchange of type I or
type III mediators leads to $\Delta L=0$ K\"ahler operators of the
form $|H_2 L|^2$, whereas the type II mediators induce both
$\Delta L=0$ and $\Delta L=2$ operators.
The type II seesaw mechanism is realized through the exchange of
$SU(2)_W$ triplet states $T$ and
$\bar{T}$ in a vector-like
$SU(2)_W\times U(1)_Y$ representation, $T\sim (3,1)$, $\bar{T}
\sim (3,-1)$. The relevant superpotential is:
$W \supset \frac{1}{\sqrt{2}}\bY^{ij}_{T} L_i T L_j
+ \frac{1}{\sqrt{2}}\la_1 H_1 T H_1 + \frac{1}{\sqrt{2}} \la_2 H_2
\bar{T} H_2 + M_T T \bar{T}\,,$
where $\bY^{ij}_{T}$ is a  $3 \times 3$ symmetric matrix,
$\la_{1,2}$ are dimensionless couplings and $M_T$ is the (SUSY)
triplet mass. By integrating out the triplets one obtains
$\Delta L=2$ effective operators of dimension  $d=5$,
\ie \, $W_{\rm eff} \supset \frac{\la_2}{2M_T}
\bY^{ij}_{T} (L_i H_2) (L_j H_2)$, and  $d=6$,
\ie \, $K_{\rm eff} \supset \frac{\la^\ast_1}{2 |M_T|^2} \bY^{ij}_T
(H^\dagger_1 L_i )(H^\dagger_1 L_j ) + h.c.$, both of which
can generate neutrino masses. The former operator is usually
leading, but we assume it to be strongly suppressed (absent) by a
very small (vanishing) value of $\la_2$, which can be
justified by symmetry arguments. In this case the leading
$\Delta L=2$ operator is the K\"ahler one, which can be
matched to the SUSY part of eq.~(\ref{K2})
through the identification: $\bkappa = \la_1^* \bY_T$ and
$M^2=|M_T|^2$. The resulting contribution to neutrino masses is
the tree-level term $\bmm^{(\kappa)}_\nu$ of eq.~(\ref{mnuk}).

One can now ask how the SUSY-breaking operators of
eq.~(\ref{K2}) arise in the type II seesaw
framework. The answer to this question depends on the ordering of the SUSY-breaking
mediation scale $M_S$ and the triplet mass $M_T$: $M_S <
M_T$; $M_S > M_T$ or $M_S = M_T$.\\
\indent $\bullet \,M_S < M_T$. If SUSY breaking mediation occurs at $M_S < M_T$
via a messenger sector coupled to the MSSM
through gauge interactions only (pure gauge mediation), then SUSY-breaking gaugino and
(flavour blind) sfermion masses arise at
$M_S$ through loop diagrams, while trilinear terms are driven below $M_S$ by gaugino mass terms in the RGEs.
As for our $\Delta L=2$ operators of
eq.~(\ref{K2}), only the SUSY one with
coefficient $\bkappa$ is present above $M_S$ (it is generated at
$M_T$), whereas the SUSY-breaking ones receive finite two-loop contributions
at $M_S$, and logarithmic ones below $M_S$ through RGEs. The dominant source
of neutrino masses is, generically, the SUSY contribution $\bmm^{(\kappa)}_\nu$ of
eq.~(\ref{mnuk}).

$\bullet\, M_S > M_T$. Suppose that SUSY-breaking terms are
generated at $M_S>M_T$ through, \eg, gravity or gauge
mediation. This means that all the MSSM and triplet
fields have SUSY-breaking mass parameters at $M_T$.
In this case, the tree-level decoupling of the triplets
generates all the $\Delta L=2$ effective operators
of eq.~(\ref{K2}), with coefficients $\bkappa = \la_1^* \bY_T$,
$\bB_\kappa = \la_1^* (\bY_T B_T - \bA_T)$,
${\tilde \bB}_\kappa = (\la_1^* B_T^* -A_1^*)
\bY_T$,
$\bC_\kappa = (\la_1^* B_T^*
-A_1^*)  (\bY_T B_T - \bA_T) - \la_1^* \bY_T m^2_{\bar{T}}$.

\section{$\bm{M_S=M_T}$: Seesaw mediators as SUSY-breaking messengers}
\vspace*{0.2cm}
The $M_S=M_T$ scenario is obtained by identifying $T$ and $\bar{T}$
as being SUSY-breaking mediators. They are embedded in a messenger sector which (in
order to generate the gluino mass) should also include coloured
fields. We also require that perturbative unification of
gauge couplings be preserved and that all messenger masses be of
the same order. This implies that the messenger sector should have
a common total Dynkin index $N$ for each subgroup of $SU(3)_C
\times SU(2)_W \times U(1)_Y$.
Since the $T+\bar{T}$ pair has $SU(2)$ index $N_2=4$, we are
constrained to $N \geq 4$. The minimal case ($N=4$) can be built by
adding a pair of $SU(3)_C$ triplets $(3,1,-1/3)+(\bar{3},1,+1/3)$
and an $SU(3)_C$ adjoint $(8,1,0)$ to $T$ and $\bar{T}$.

The MSSM SUSY-breaking parameters are generated at the quantum level
by a messenger sector of the type described above, coupled
to the MSSM fields through both gauge and Yukawa interactions~\cite{JR}. At
the one-loop level, the gaugino masses $M_a$, the Higgs $B$-term
$B_H$ and the trilinear terms $\bA_x$ are:
%%%%%%%%%%%%
$M_a =  -\frac{N B_T }{16 \pi^2}\, g_a^2$,
$B_H  = \frac{3 B_T }{16 \pi^2}\, |\la_1|^2$
${\bA}_e =  \frac{3 B_T}{16 \pi^2} \bY_e (\bY^\dagger_T \bY_T
+|\la_1|^2 )$, ${\bA}_d = \frac{ 3 B_T}{16 \pi^2} \bY_d
|\la_1|^2$ and ${\bA}_u = 0$,
where $g_1^2=(5/3)g'^2$ and $g_2^2=g^2$. Non-vanishing ${\cal
O}(B_T^2)$ contributions for the squared scalar masses arise at
the two-loop level. For sleptons these are:
\bea \label{mlsoft} \!\!\!\!\bmm^2_{\tl}& =& \left(\frac{|B_T|}{16
\pi^2}\right)^2 \left[ N \left( \frac{3}{10} g^4_1 + \frac{3 }{2}
g^4_2 \right) - \left(\frac{27}{5} g^2_1 +21
g^2_2\right)\!\bY^\dagger_T\bY_T + 3 |\la_1|^2 (\bY^\dagger_T
\bY_T- \bY^\dagger_e \bY_e)
\right. \no \\
&& \phantom{xxxxxxx} \left. + 3\, \bY^\dagger_T (\bY^\dagger_e
\bY_e)^T \bY_T + 18 \, (\bY^\dagger_T\bY_T)^2 + 3 \,
\bY^\dagger_T\bY_T {\rm Tr}(\bY^\dagger_T\bY_T)
\right]\,, \\
\!\! \bmm^2_{\te}& =& \left(\frac{|B_T|}{16 \pi^2}\right)^2
\left[N \left( \frac{6 }{5} g^4_1 \right) - 6 \, \bY_e(
\bY^\dagger_T\bY_T+
 |\lambda_1|^2 )\bY^\dagger_e \right]\,.
\eea
Notice that the
flavour structures of $\bA_e$, $\bmm^2_\tl$ and $\bmm^2_{\te}$ are
controlled by $\bY_T$ and $\bY_e$, which in turn are determined
by the low-energy lepton masses and mixing angles. Such
minimal LFV properties are a characteristic feature of the SUSY
type II seesaw \cite{JR,AR}. Our scenario is a flavoured variant
of gauge mediation and possesses the
property of minimal flavour violation in both the quark
and lepton sectors.

Upon decoupling the triplets, the MSSM SUSY-breaking masses are
generated through finite radiative effects, while the $\Delta L=2$
SUSY-breaking parameters $\bB_\kappa$, ${\tilde \bB}_\kappa$ and
$\bC_\kappa$ arise at the tree level. The latter have a very
simple form: $\bB_\kappa = B_T \, \bkappa$, ${\tilde
\bB}_\kappa = B_T^* \, \bkappa$ and $\bC_\kappa
=|B_T|^2 \, \bkappa$, where $ \bkappa =  \la_1^* \bY_T$.
Eqs.~(\ref{mnuk}) and (\ref{mnubtk}) imply that the tree-level
contribution to $\bmm_\nu$ is:
\be \label{mnusum}
\bmm_\nu = \bkappa \, ( B_T^* +  2\, \mu \tan\beta) \cos^2 \!
\beta \; \frac{ v^2}{|M_T|^2} \, .
\ee
Notice that the SUSY-breaking parameter $B_T$ acts as a common
source for both sparticle and neutrino masses.

This scenario has a small number of free parameters, namely
$M_T$, $B_T$, $\lambda_1$ and the messenger index $N$. Once these
are fixed, the remaining parameters $\bY_T$, $\tan\beta$ and $\mu$
are determined by the low-energy neutrino data and by requiring
proper EWSB. Concerning neutrino data, we recall that the neutrino
mass matrix $\bmm_\nu$ is related to the low-energy observables as
$\bmm_\nu = \bU^* \bmm^{D }_\nu \bU^\dagger$, where $\bmm^{D }_\nu
= {\rm diag}(m_1,m_2,m_3)$, $m_a$ are the neutrino masses and
$\bU$ is the lepton mixing matrix.
%
%%%%%%%%%%%%%%%%%  fig. 1 %%%%%%%%%%%%%%%%%%%%%%%%%%%%%%%%%%%%%%%%%%%%
\begin{figure*}[t] \begin{center} \begin{tabular}{cc}
\includegraphics[width=10.0cm]{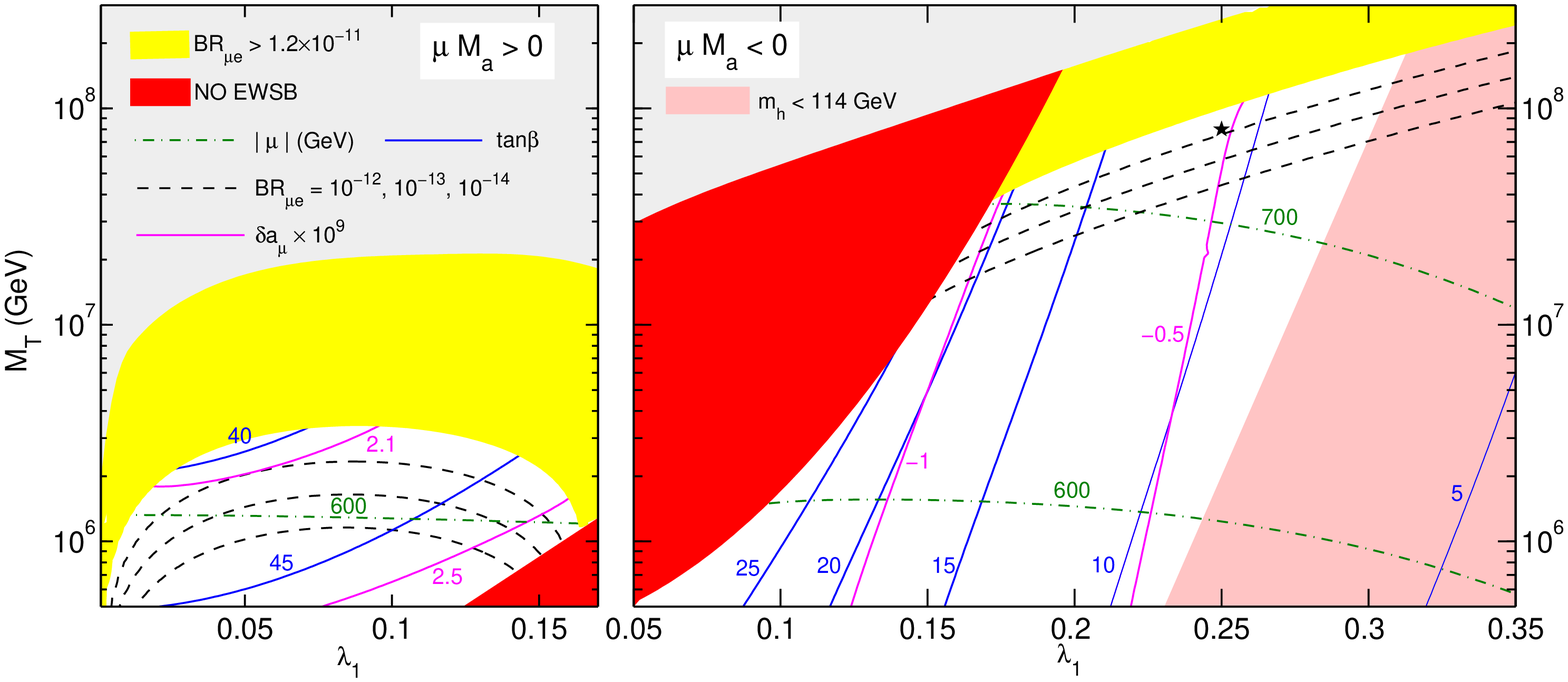} & \includegraphics[width=5.3cm]{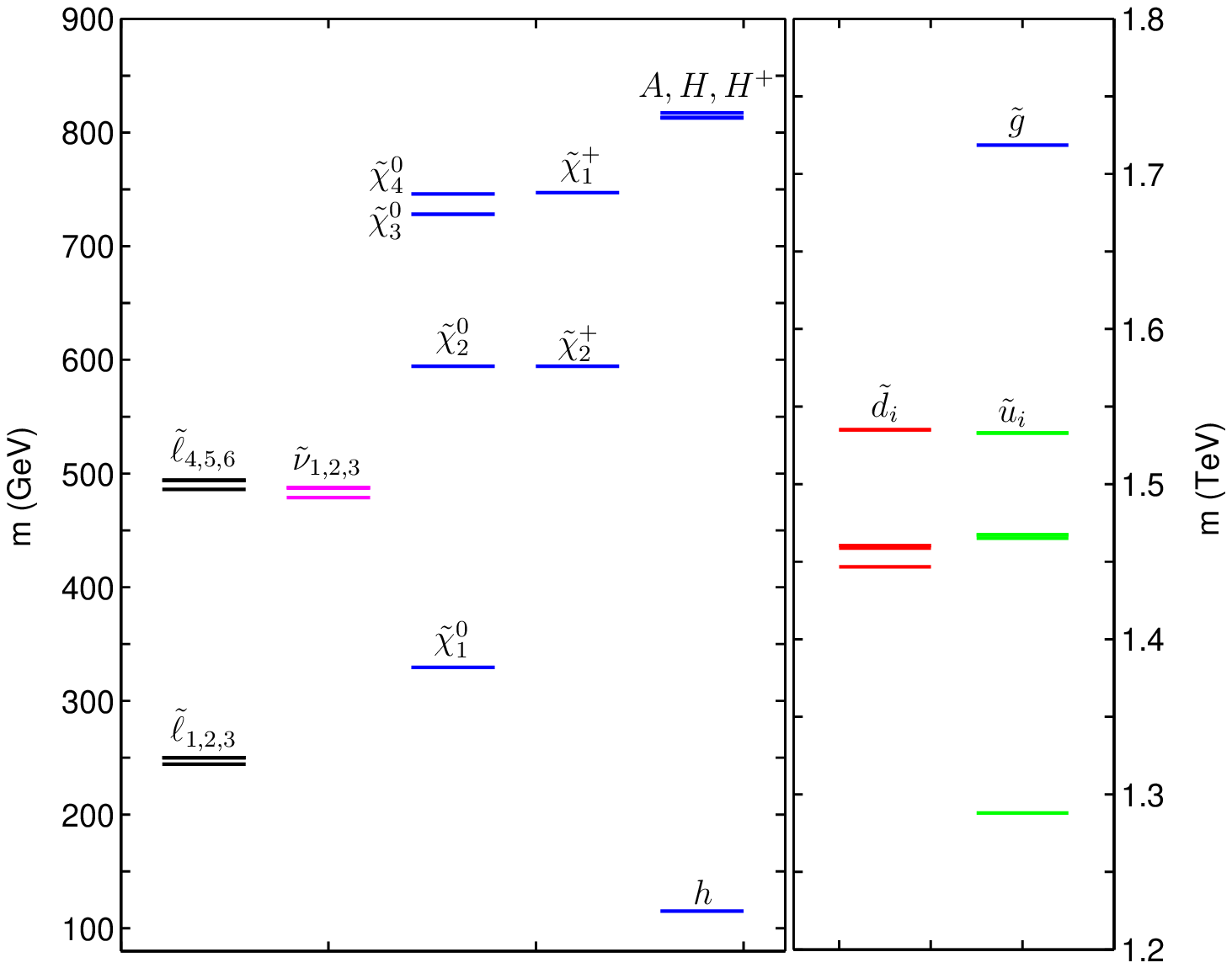}
\end{tabular}
\caption{Plots of the $N=4$ model for
$B_T=60~{\rm TeV}$ and normally ordered neutrino spectrum with
$0=m_1<m_2 \ll m_3$ and $s_{13}=0$. {\em Left panels}: The
$(\la_1, M_T)$ parameter space for $\mu M_a >0$ (left) and  $\mu
M_a <0$ (right). The white region is the allowed one.
The dashed lines correspond to $\BR(\mu\to e \ga)=10^{-12},
10^{-13}, 10^{-14}$ (from top to bottom.). {\em Right
panel}: Sparticle and Higgs spectrum for $M_T=8 \times 10^7 \,
\gev$ and $\la_1=0.25$ (point $\star$ of the parameter space).}
\label{f5} \end{center} \end{figure*}
%%%%%%%%%%%%%%%%%%%%%%%%%%%%%%%%%%%%%%%%%%%%%%%%%%%%%%%%%%%%%%
%%%%%%%%%%%%%%%%%%%%%%%%%%%%%%%%%%%%%%%%%%%%%%%%%%%%%%%%%%
Some representative numerical results for the $N=4$ model are
shown in  Fig.~\ref{f5}, for $B_T=60~{\rm TeV}$ and a normally ordered
neutrino spectrum, with $0=m_1<m_2 \ll m_3$ and $s_{13}=0$. In the
left part of Fig.~\ref{f5} we show two plots of the $(\la_1,M_T)$
parameter space, including contours of $\tan\beta$ and $\mu$
(extracted by imposing EWSB). The left (right) panel corresponds
to solutions of the EWSB conditions with $\mu M_a >0$ ($\mu M_a
<0$ ). The main phenomenological constraints come from the LFV
decay $\mu \rightarrow e \gamma$ and the lightest Higgs mass.
In the right panel of Fig.~\ref{f5} we show the sparticle and
Higgs spectrum for $M_T=8 \times 10^7 \, \gev$ and
$\lambda_1=0.25$ (which corresponds to $\tan\beta \simeq 11$),
again for $B_T=60~{\rm TeV}$. The Higgs sector is in the decoupling
regime, since the states $A,H$ and $H^+$ are much
heavier than $h$. Gluino and squarks are the heaviest sparticles
and the lightest of them is ${\tilde t}_1$ (which is mainly
${\tilde t}_R$). In the electroweak sector, the heaviest chargino
and neutralinos (${\tilde \chi}^+_2$, ${\tilde \chi}^0_{3,4}$) are
mainly Higgsino-like, while ${\tilde \chi}^+_1$ and ${\tilde
\chi}^0_{2}$ are mostly Wino-like. The
(mainly left-handed) sleptons ${\tilde \ell}_{4,5,6}$ and the
sneutrinos ${\tilde \nu}_{1,2,3}$ are somewhat lighter than those
states, and the Bino-like neutralino ${\tilde \chi}^0_{1}$ is even
lighter.
Finally, the lightest MSSM sparticles are the (mainly
right-handed) sleptons ${\tilde \ell}_{1,2,3}$, as generically
occurs in gauge mediated models with messenger index $N
>1$ and not too large mediation scale. The slepton
${\tilde \ell}_{1}$ (which is mainly $\tilde{\tau}_R$) is the
next-to-lightest SUSY particle (NLSP), while the gravitino
${\tilde G}$ is the lightest one (LSP).

The scenario described above can be tested at current and future
colliders. For instance, if a $p p$ collision at the LHC produces
two squarks, each of them can decay through well known chains,
such as $\tilde{q}_R \rightarrow q {\tilde \chi}^0_1 \rightarrow q
\tau {\tilde \ell}_1$, or $\tilde{q}_L \rightarrow q {\tilde
\chi}^0_2
 \rightarrow q \ell  {\tilde \ell} \rightarrow
q  \ell^+ \ell^- {\tilde \chi}^0_1  \rightarrow q  \ell^+ \ell^-
\tau {\tilde \ell}_1$, or similar ones with charginos and/or
sneutrinos (and neutrinos). Hence, in general, the final state of
such a collision contains SM particles and two NLSPs
${\tilde \ell}_{1}$, which eventually decay to $\tau {\tilde G}$
with rate $\Gamma= m^5_{{\tilde \ell}_1}/(16 \pi F^2)$. The latter
decays can occur either promptly, or at a displaced vertex, or even
outside the main detector.

Since LFV is intrinsically present in our framework, LFV
processes are a crucial tool to discriminate our model from pure
gauge mediation ones. For moderate $\tan\beta$, the leading
LFV structure $\bY^\dagger_T \bY_T$, which appears in
$\bmm^2_{\tl}$ (and $\bA_e$), can be related to the neutrino
parameters as
\beq\label{rel1} (\bmm^2_{\tl})_{ij}    \propto
B_T^2 (\bY^{\dagger}_T \bY_T)_{ij}  \propto \left(\frac{M_T^2
\tan^2 \! \beta }{\la_1}\right)^2
 \left[\bV (\bmm^{D }_\nu)^2 \bV^\dagger\right]_{ij} \propto
\tan^5 \! \beta \, M_T^4
 \left[\bV (\bmm^{D }_\nu)^2 \bV^\dagger\right]_{ij}\,.
\eeq
LFV signals can therefore appear either at high-energy colliders
or in low-energy processes. Concerning the former possibility, LFV
could show up at the LHC in, \eg , neutralino decays, such as
${\tilde \chi}^0_2 \rightarrow \ell_i^{\pm} {\tilde \ell}_a^{\mp}
\rightarrow \ell_i^{\pm} \ell_j^{\mp} {\tilde \chi}^0_1$ with $i
\not= j$, followed by the flavour-conserving decay ${\tilde
\chi}^0_1\rightarrow \tau^+ \tau^- {\tilde G}$ (or ${\tilde
\chi}^0_1  \rightarrow {\tilde \ell}_1^{\pm} \tau^{\mp}$ if the
NLSP is long-lived).

As for low-energy LFV processes, let us focus on the radiative
decays $\ell_i \to \ell_j \ga $. By using eq.~(\ref{rel1}), we can
infer that \be\BR(\ell_i \to \ell_j \ga) \propto
\frac{|(\bmm^2_{\tl})_{ij}|^2} { \tm^8 }
 \tan^2\!\beta
\propto \left(\frac{M_T}{B_T}\right)^{\!8 }\!(\tan\beta)^{12}
\left|\left[\bV (\bmm^{D }_\nu)^2
\bV^\dagger\right]_{ij}\right|^2\,,\ee
where the flavour-dependence is
determined by the low-energy neutrino parameters only \cite{AR, JR}.
If we take ratios of $\BR$s, we have that,
for a normal neutrino mass spectrum, $\BR(\tau \to \mu \ga)/\BR(\mu \to e \ga)\simeq 400\,(2)\,[3]$ and
$\BR(\tau \to e \ga)/\BR(\mu \to e \ga)\simeq 0.2\,(0.1)\,[0.3]$ if
$s_{13}=0\,(s_{13}=0.2,\delta=0)\,[s_{13}=0.2,\delta=\pi]$.
These approximate results hold for small or moderate
$\tan\beta$. Special features may emerge
for large $\tan\beta$~\cite{FJ}.

\begin{figure}[t] \begin{center} \begin{tabular}{ll}
\includegraphics[width=7.3cm]{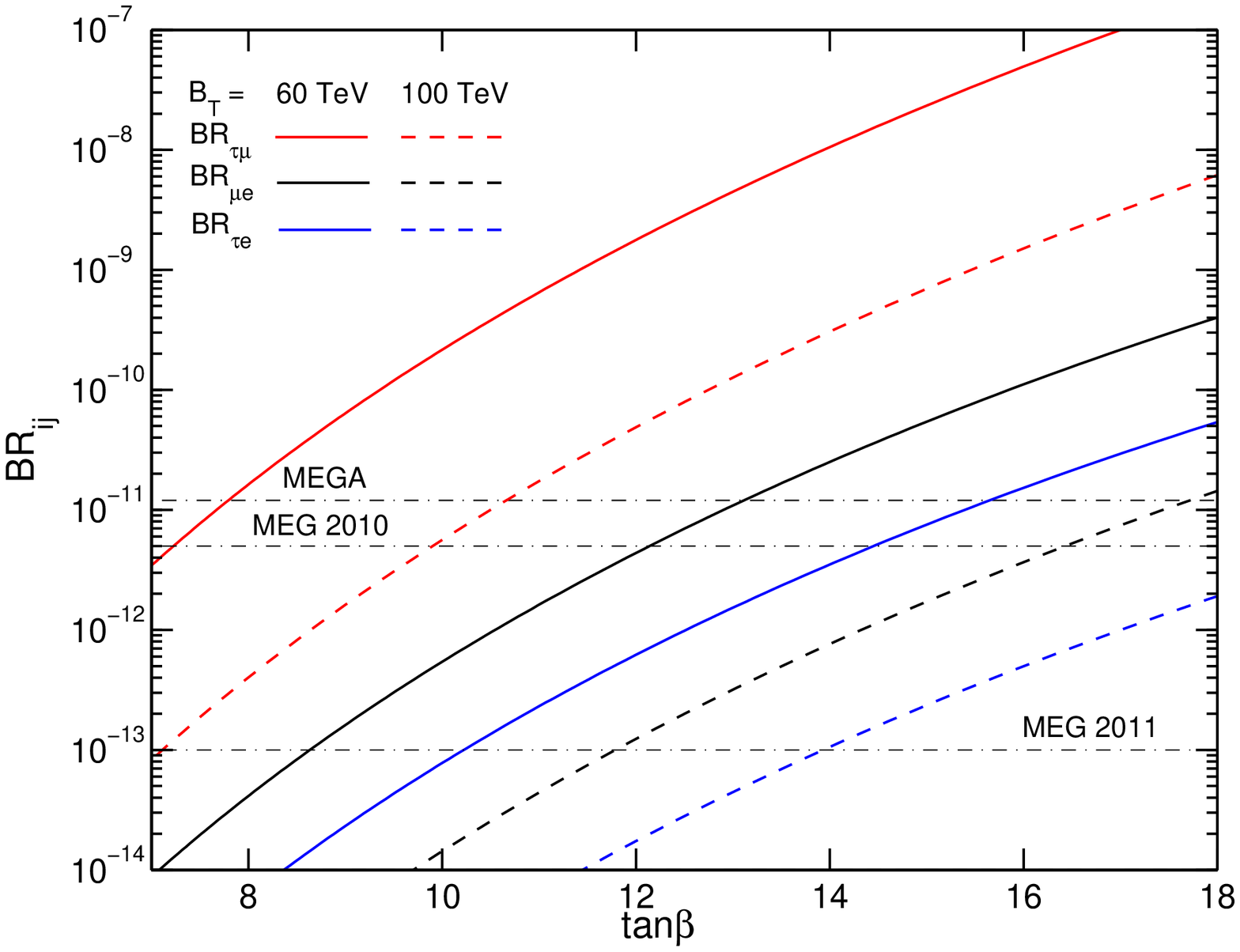} &
\includegraphics[width=7.6cm]{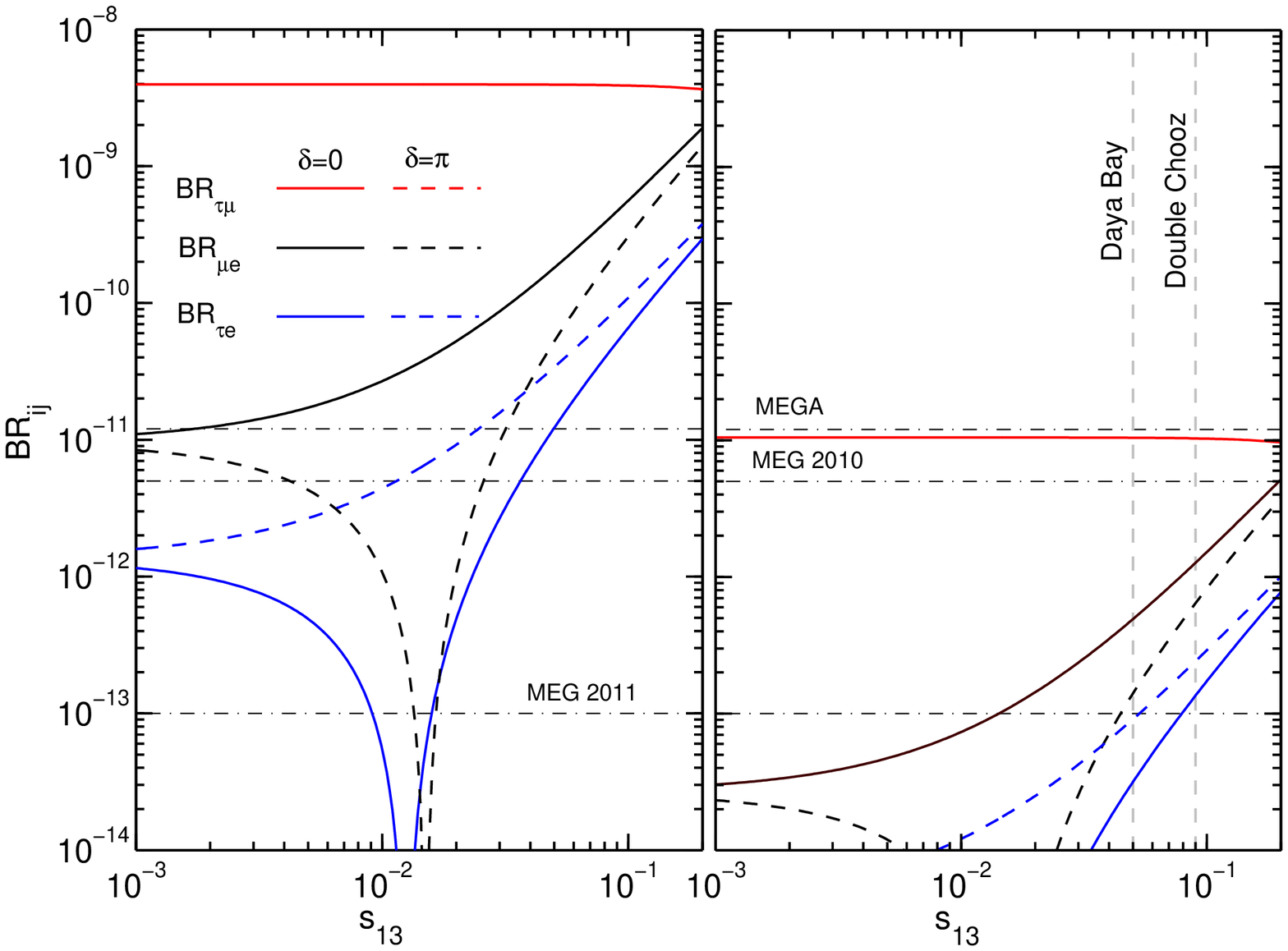} \end{tabular}
\caption{Plots of $\BR(\ell_i \to \ell_j \gamma)$ for $M_T =
8\times 10^7\,{\rm GeV}$. {\it Left panel}: The BRs as a function
of $\tan\beta$ for $B_T= 60\,{\rm TeV}$ (solid lines) or $B_T=
100\,{\rm TeV}$ (dashed lines), with $s_{13}=0$. {\it Right
panel}: The BRs as a function of $s_{13}$ for $B_T= 60\,{\rm
TeV}$, with $\tan\beta = 13 \, (8)$ in the first (second)
subpanel. For $\BR(\mu \to e \gamma)$ and $\BR(\tau \to e
\gamma)$, the solid (dashed) curves correspond to $\delta = 0 \,
(\pi)$ assuming a normally-ordered neutrino spectrum.} \label{f6}
\end{center} \end{figure}

In the left panel of
Fig.~\ref{f6} the three $\BR(\ell_i \to \ell_j \ga)$ are shown as
a function of $\tan\beta$, taking either $B_T= 60\,{\rm TeV}$
(solid lines) or $B_T= 100\,{\rm TeV}$ (dashed lines), for $M_T =
8\times 10^7\,{\rm GeV}$, normally ordered
neutrino spectrum and $s_{13}=0$. In both examples
$\BR(\mu \to e \ga)$ can be tested at the MEG
experiment for suitable ranges of  $\tan\beta$.
If $\BR(\mu \to e \ga)$ is close
to its present bound, $\BR(\tau \to \mu \ga)$ is above $10^{-9}$,
within the reach of future Super Flavour Factories.

%%%%%%%%%%%%%%%%%%%%%%%%%%%%%%%%%%%%%%%%%%%%%%%%%%%%%%%%555
%%%%%%%%%%%%%%%%%%%%%%%%%%%%%%%%%%%%%%%%%%%%%%%%%%%%%%%%%%

The double panel on the right of Fig.~\ref{f6} illustrates the
dependence of the BRs on the least known neutrino parameters,
namely $s_{13}$ and $\delta$, for $B_T= 60\,{\rm TeV}$, $M_T =
8\times 10^7\,{\rm GeV}$ and two values of $\tan\beta$.
Regarding $\BR(\mu \to e \gamma)$ and
$\BR(\tau \to e \gamma)$, the solid (dashed) curves correspond to
$\delta = 0 \, (\pi)$, while the region between such curves is
spanned by intermediate values of $\delta$. The dependence of
$\BR(\tau \to \mu \gamma)$ on $s_{13}$ and $\delta$ is negligible.
The first subpanel corresponds to a scenario which could be tested
very soon at MEG through the search of $\mu \to e \gamma$, if
$s_{13} \ll 0.01$. Notice that $\BR(\tau \to \mu \gamma)$ in this
example is around $4 \times 10^{-9}$, within the reach of future
Super Flavour Factories, while $\tau \to e \gamma$
would be unobservable because $\BR(\tau \to e \gamma)\sim
10^{-12}$. For $s_{13} \sim 0.01$, $\BR(\mu \to e \gamma)$ and
$\BR(\tau \to e \gamma)$ can be either enhanced or suppressed
since, depending on the value of $\delta$, a cancellation can
occur in the LFV quantity $\left[\bV (\bmm^{D }_\nu)^2
\bV^\dagger\right]_{ij}$ \cite{JR,FJ}. The cancellation takes place in $\BR(\mu \to e
\gamma)$ [$\BR(\tau \to e \gamma)$] for $\delta = \pi \, (0)$ in
the case of normal ordering, while the opposite occurs for
inverted ordering. If $\BR(\mu \to e \gamma)$ is suppressed by
that cancellation mechanism, only $\tau \to \mu \gamma$ can be
observed. In such a case,  we can even obtain values of $\BR(\tau
\to \mu \gamma)$ above $ 10^{-8}$
by slightly changing the model parameters. In the case of partial cancellations, $\mu \to
e \gamma$ could be still probed by MEG for values of $s_{13}$ up
to about 0.03, which are in the potential reach of future Neutrino
Factories. The second subpanel shows an alternative
possibility, in which LFV $\tau$ decays are unobservable, whereas
$\BR(\mu \to e \gamma)$ lies in the range $10^{-13} - 5 \times
10^{-12}$ if $0.05 \sim s_{13} < 0.2$. Those values of $\BR(\mu
\to e \gamma)$ should be probed by MEG next year, while the
indicated range of $s_{13}$ is within the sensitivity of the
present and incoming neutrino experiments.
This example shows the importance of the interplay between LFV
searches and neutrino oscillation experiments.

%%%%%%%%%%%%%%%%%%%%%%%%%%%%%%%%%%%%%%%%%%%%%%%%%%%%%%%%%%
\vspace*{0.2cm}
In conclusion, we have summarized the investigation of
Ref.~\cite{Brignole:2010nh} on an interesting alternative neutrino mass
mechanism, which relies on the $d
 = 6$, $\Delta L=2$ K\"{a}hler
operator $(H^\dagger_1 L)^2/M^2$. We have presented both
a general effective-theory description and
explicit realizations in the context of the type II
seesaw mechanism. If the seesaw mediators are also identified with
SUSY-breaking messengers, strong correlations arise between neutrino
parameters, sparticle and Higgs masses, as well as LFV processes.

\section*{References}

%--------------------------------------------------------------

\end{document}